# Development of an Adaptive Sliding Mode Controller using Neural Networks for Trajectory Tracking of a Cylindrical Manipulator

TieuNien Le[1,2], VanCuong Pham[1], NgocSon Vu[1,*]

[1]Faculty of Electrical Engineering, Hanoi university of Industry, Hanoi, Vietnam
[2]Falcuty of Electrical and Electronic Engineering, East Asia University of Technology, Bac Ninh, Vietnam

**ABSTRACT:** Cylindrical manipulators are extensively used in industrial automation, especially in emerging technologies like 3D printing, which represents a significant future trend. However, controlling the trajectory of nonlinear models with system uncertainties remains a critical challenge, often leading to reduced accuracy and reliability. To address this, the study develops an Adaptive Sliding Mode Controller (ASMC) integrated with Neural Networks (NNs) to improve trajectory tracking for cylindrical manipulators. The ASMC leverages the robustness of sliding mode control and the adaptability of neural networks to handle uncertainties and dynamic variations effectively. Simulation results validate that the proposed ASMC-NN achieves high trajectory tracking accuracy, fast response time, and enhanced reliability, making it a promising solution for applications in 3D printing and beyond.

## I. INTRODUCTION

Cylindrical manipulators play a critical role in industrial automation due to their flexibility and high precision, particularly in cutting-edge applications such as 3D printing, a future-defining technology. These manipulators must perform accurate trajectory tracking under dynamic and uncertain conditions, making robust and adaptive control systems crucial for ensuring operational efficiency. However, nonlinear dynamics, external disturbances, and system uncertainties pose significant challenges, often resulting in reduced accuracy and reliability in real-world applications. Addressing these challenges is an urgent requirement to enhance the performance of control systems. [1-3]

Various control strategies have been proposed to improve the trajectory tracking capabilities of robotic manipulators. Traditional Proportional-Integral-Derivative (PID) controllers remain popular due to their simplicity and ease of implementation. However, they are less effective in managing complex nonlinear systems and uncertainties [4-6]. A study in [4] focused on PID tracking control of robot manipulators, aiming to improve precision and efficiency. Researchers in [5] developed a non-linear PID controller for trajectory tracking of a differential drive mobile robot to enhance system responsiveness. Additionally, [6] proposed an efficient PID tracking control method for robotic manipulators driven by compliant actuators to boost performance and accuracy. These studies have significantly contributed to enhancing control system performance and precision but require further optimization to effectively handle nonlinearities and complex dynamic scenarios.

Sliding Mode Controllers (SMC) are recognized for their robustness against disturbances and uncertainties [7-9]. In [7], robust sliding mode control was developed for robot manipulators to improve stability and robustness under various disturbances. Research in [8] combined PD with sliding mode control for trajectory tracking in robotic systems, aiming to enhance precision and response times. Furthermore, [9] explored trajectory planning and second-order sliding mode motion/interaction control for robotic manipulators operating in unknown environments, focusing on improving adaptability and interaction capabilities. These studies highlight significant advancements in sliding mode control techniques. However, the chattering phenomenon limits their performance and can cause mechanical wear [10]. To overcome these limitations, recent studies have focused on integrating artificial intelligence technologies, particularly Neural Networks (NNs), to approximate system uncertainties and enhance adaptability. While hybrid methods have shown promise, achieving a balance between robustness and adaptability remains a significant challenge [11-17]. In [11], fuzzy-neural-network inherited sliding-mode control was developed for robot manipulators, incorporating actuator dynamics to enhance control performance. In [12], a neural network-based sliding mode adaptive control method was proposed to improve the adaptability and precision of robot manipulators. The study in [13] focused on robust adaptive sliding mode neural networks control for industrial robot manipulators, aiming to enhance system robustness and stability. Additionally, [14] explored the use of fuzzy wavelet neural networks in robust adaptive sliding mode control for industrial robots, targeting improved performance under varying conditions. Further, [15] introduced recurrent fuzzy wavelet neural networks for robust adaptive sliding mode control of industrial robot manipulators, emphasizing enhanced learning and control capabilities. Lastly, in [16], neural-network-based terminal sliding-mode control was examined, including actuator dynamics to achieve better control accuracy and responsiveness. These studies demonstrate the significant advancements in robotic control through the integration of Sliding Mode Control and neural networks, leading to enhanced precision, robustness, and adaptability of robotic manipulators in complex and dynamic environments [17].

This study addresses the challenges of cylindrical manipulators by developing a Sliding Mode Controller (SMC) integrated with Neural Networks (NNs). The proposed controller combines the robustness of sliding mode control to handle disturbances and uncertainties with the adaptability of neural networks, which estimate system variations and adjust the control signals accordingly. This framework is applied to the trajectory tracking problem in 3D printing, a rapidly growing technology with stringent performance demands. Simulation results demonstrate that the proposed controller achieves high trajectory tracking accuracy, fast response times, and improved reliability, paving the way for its application in 3D printing and other industrial fields.

In summary, this study emphasizes the importance of robust control systems for cylindrical manipulators and introduces an innovative ASMC-NN approach to address key trajectory tracking challenges. The findings lay a foundation for enhancing the precision and reliability of cylindrical manipulators in advanced automation applications.

## II. MODELING OF CYLINDRICAL MANIPULATOR

To develop an effective control strategy for a cylindrical manipulator, a precise mathematical model is essential. This model serves as the basis for designing the Adaptive Sliding Mode Controller, enabling accurate trajectory tracking despite the presence of system uncertainties.

A cylindrical manipulator typically consists of three main degrees of freedom:

+ Rotational Motion (Joint 1): Rotation around the base, allowing the manipulator to access a circular workspace.

+ Vertical Linear Motion (Joint 2): Linear movement along the vertical axis, facilitating height adjustments.

+ Radial Linear Motion (Joint 3): Linear movement along the radial axis, enabling the end-effector to reach varying distances from the base.

This configuration makes cylindrical manipulators particularly suited for tasks requiring precise positioning within a constrained workspace, such as 3D printing.

The dynamics of the cylindrical manipulator are derived using the Lagrangian formulation, which describes the system's total energy:

$$L(q,\dot{q}) = K(q,\dot{q}) - P(q) \qquad (1)$$

Where: $L(q,\dot{q})$ is the Lagrangian, representing the system's total energy, $K(q,\dot{q})$ is the kinetic energy, $P(q)$ is the potential energy.

The Lagrangian approach leads to the following equation of motion:

$$M(q)\ddot{q} + C(q,\dot{q})\dot{q} + F(q)q + G(q) = \tau - f_{ext} \qquad (2)$$

where: $(q,\dot{q},\ddot{q}) \in R^{n \times 1}$ are joint position, velocity, and acceleration vectors. $M(q) \in R^{n \times n}$ is the symmetric and positive definite inertia matrix.. $C(q,\dot{q}) \in R^{n \times n}$ represents the Coriolis and centripetal forces. $G(q) \in R^{n \times 1}$ is the gravity vector. $F(q)$ accounts for friction forces. $f_{ext} \in R^{n \times 1}$ represents external disturbance forces. $\tau \in R^{n \times 1}$ is the joints torque input vector.

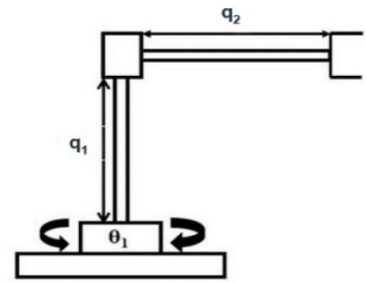

**Figure 1. The cylindrical manipulator**

For a three-joint cylindrical manipulator, the dynamics are expressed in matrix form:

$$\begin{bmatrix}\tau_1\\\tau_2\\\tau_3\end{bmatrix} = \begin{bmatrix}A_{11} & A_{12} & A_{13}\\A_{21} & A_{22} & A_{23}\\A_{31} & A_{32} & A_{33}\end{bmatrix} \times \begin{bmatrix}\ddot{\theta}_1\\\ddot{q}_2\\\ddot{q}_3\end{bmatrix} + \begin{bmatrix}B_{11} & B_{12} & B_{13}\\B_{21} & B_{22} & B_{23}\\B_{31} & B_{32} & B_{33}\end{bmatrix} \times \begin{bmatrix}\dot{\theta}_1^2\\\dot{q}_2^2\\\dot{q}_3^2\end{bmatrix} + \begin{bmatrix}C_{11} & C_{12} & C_{13}\\C_{21} & C_{22} & C_{23}\\C_{31} & C_{32} & C_{33}\end{bmatrix} \times \begin{bmatrix}\dot{\theta}_1\dot{q}_2\\\dot{\theta}_1\dot{q}_3\\\dot{q}_2\dot{q}_3\end{bmatrix} + \begin{bmatrix}D_1\\D_2\\D_3\end{bmatrix}$$

where:
$A_{11} = (4m_1 \sin\theta_1 - 4m_2 \cos\theta_1)q_3 + I_3$
$A_{13} = (m_1 + m_2)(\sin\theta_1 \cos\theta_1)q_3$
$A_{22} = m_3; A_{31} = m_1 \sin\theta_1 \cos\theta_1$
$A_{33} = 2(m_1 \sin\theta_1 + m_2 \cos\theta_1); A_{12} = A_{21} = A_{23} = A_{32} = 0;$
$B_{11} = (m_1 \sin\theta_1 - 4m_2 \cos\theta_1)q_3$
$B_{13} = -m_1 \cos\theta_1 + m_2 \sin\theta$
$B_{31} = 2q_3(m_1 \sin\theta_1 - m_2 \cos\theta_1)$ ;

$B_{12} = B_{21} = B_{22} = B_{23} = B_{32} = B_{33} = 0$
$C_{12} = -(m_1 + m_2)(\sin\theta_1 \cos\theta_1)q_3$
$C_{32} = -(m_1 + m_2)(\sin\theta_1 \cos\theta_1)$
$C_{11} = C_{13} = B_{21} = C_{22} = C_{23} = C_{31} = C_{33} = 0$
$D_2 = g(m_2+m_3); D_1 = D_3 = 0$

Where: $m_i$ are the mass of joint and $l_i$ are the length of joint of Cylindrical Manipulator, respectively, $I_3 = 1$ (kg/m$^2$) is the moment of inertia of joint 3, respectively, and $g = 9.8$ (m/s$^2$) is acceleration of gravity. The values of the parameters of each joint of the manipulator are shown in Table 1.

**Table 1. Values of masses (m) and lengths (l) of each joint**

| Joint | m (kg) | l (m) |
|---|---|---|
| 1 ($\theta_1$) | 36.367405 | 0.05 |
| 2 ($q_2$) | 12.632222 | 0.79 |
| 3 ($q_3$) | 23.735183 | 0.9 |

The dynamics provide a comprehensive framework for understanding the cylindrical manipulator's motion and developing the ASMC. Future sections will focus on integrating Neural Networks to improve adaptability and address system uncertainties effectively.

## III. ADAPTIVE SLIDING MODE CONTROLLER WITH NEURAL NETWORKS (ASMC-NN)

The Adaptive Sliding Mode Controller with Neural Networks (ASMC-NN) combines the robustness of Sliding Mode Control (SMC) with the adaptive capabilities of Neural Networks (NNs) to address system uncertainties and improve trajectory tracking performance. This section details the design and mathematical formulation of the proposed controller.

* Sliding Mode Control Design

The SMC framework is designed to ensure system states converge to the desired trajectory despite disturbances and uncertainties. The system dynamics (2) are given by:
$M(q)\ddot{q} + C(q,\dot{q})\dot{q} + F(q)q + G(q) = \tau - f_{ext}$
Define the tracking error as: $e(t) = q_d(t) - q(t); \dot{e}(t) = \dot{q}_d(t) - \dot{q}(t)$
The sliding surface $s(t)$ is defined to ensure error convergence:
$$s(t) = \dot{e}(t) + \lambda e(t) \quad (3)$$
Where $\lambda > 0$ is a design parameter.
The control input $\tau$ for SMC is designed as:
$$\tau = \tau_{eq} + \tau_{sw} \quad (4)$$
+ Equivalent Control $\tau_{eq}$: Ensures the system remains on the sliding surface.
$$\tau_{eq} = M(q)\ddot{q}_d + C(q,\dot{q})\dot{q}_d + G(q) \quad (5)$$
+ Switching Control $\tau_{sw}$: Compensates for disturbances and uncertainties.
$$\tau_{sw} = -kSign(s) \quad (6)$$
Where $k > 0$ is a gain parameter, and $Sign(s)$ introduces robustness but can lead to chattering.
To reduce chattering $Sign(s)$ can be replaced with a continuous approximation such as:

$$Sign(s) \approx \frac{s}{\epsilon + |s|} \quad (7)$$
Where $\epsilon > 0$ is a small positive constant.
* Neural Network Integration
Neural Networks are introduced to approximate system uncertainties $\Delta\tau$, which are difficult to model explicitly. The control law is modified as:
$$\tau = \tau_{eq} + \tau_{sw} + \tau_{NN} \quad (8)$$
Where $\tau_{NN}$ is the output of the neural network, approximating $\Delta\tau$
A feedforward NN is used, consisting of:
  + Input Layer: $q, \dot{q}, s$
  + Hidden Layers: Nonlinear activation functions
  + Output Layer: Single output $\tau_{NN}$
The NN is trained using a backpropagation algorithm to minimize the error:
$$E = \frac{1}{2}\|\Delta\tau - \tau_{NN}\|^2 \quad (9)$$
The modified control law becomes:
$$\tau = \tau_{eq} - k\frac{s}{\epsilon+|s|} + \tau_{NN} \quad (10)$$
* Adaptive Mechanism
To improve performance, the NN weights $W$ are updated online using an adaptation law:
$$\dot{W} = \gamma\frac{\partial E}{\partial W} = \gamma s\phi(q,\dot{q},s) \quad (11)$$
Where $\gamma > 0$ is a learning rate, and $\phi$ represents the NN activation function.
Using a Lyapunov candidate function:
$$V = \frac{1}{2}s^2 + \frac{1}{2}Tr(W^TW) \quad (12)$$
The derivative of $V$ is:
$$\dot{V} = S + Tr(W^T\dot{W}) \quad (13)$$
Substituting the control law and adaptation rule ensures $\dot{V} < 0$ guaranteeing stability.

"Development of an Adaptive Sliding Mode Controller using Neural Networks for Trajectory Tracking of a Cylindrical Manipulator"

Combining SMC and NN, the unified control law is:

$$\tau = M(q)\ddot{q}_d + C(q,\dot{q})\dot{q}_d + G(q) - k\frac{s}{\epsilon+|s|} + \tau_{NN} \quad (14)$$

The adaptive mechanism dynamically adjusts NN weights to minimize approximation errors. Simulation results in the next section will demonstrate the effectiveness of ASMC-NN in trajectory tracking and robustness against disturbances.

## IV. SIMULATION RESULTS

- Constant Desired Trajectory

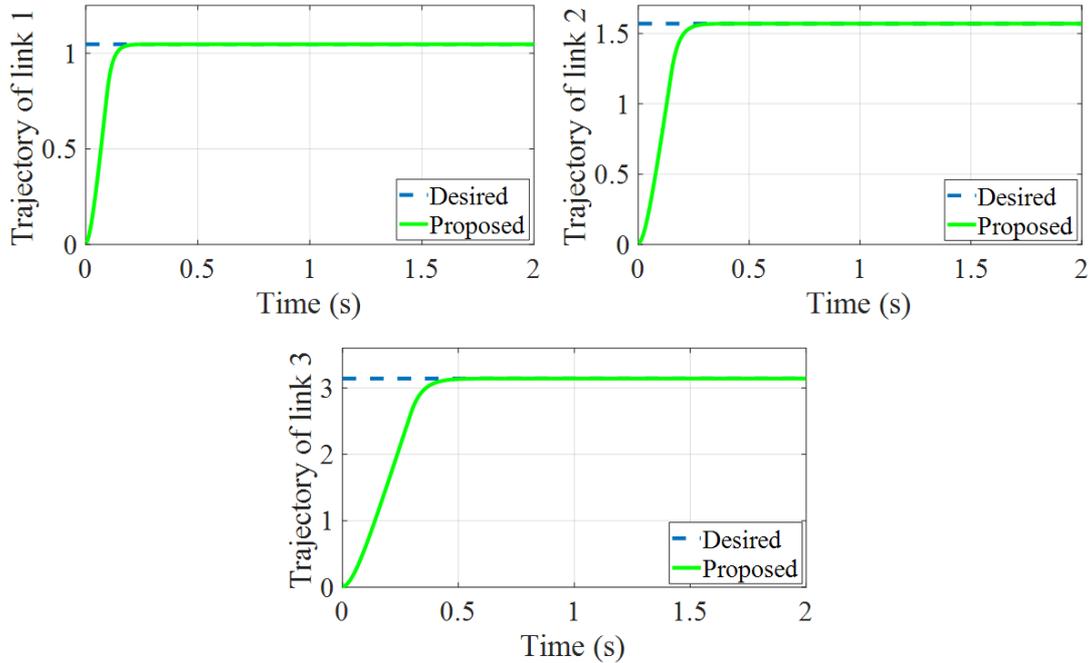

Figure 2. Trajectory responses of joints for the constant desired trajectory

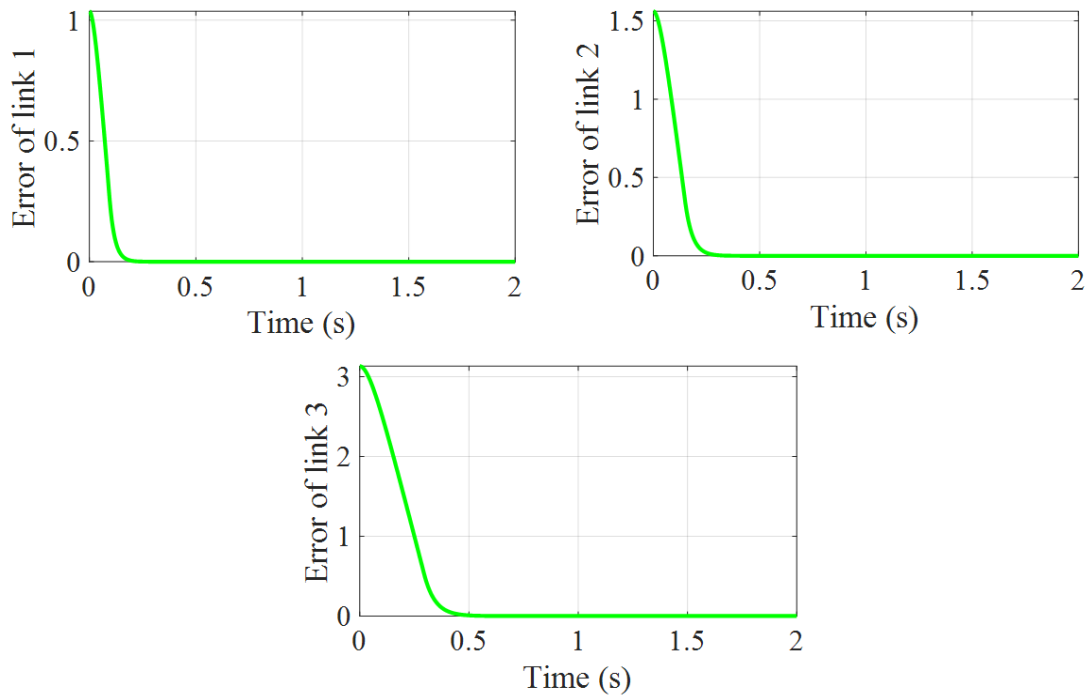

Figure 3. Tracking errors of joints for the constant desired trajectory

The desired joint values were set as: $\theta_1 = \frac{\pi}{3}, q_2 = \frac{\pi}{2}, q_3 = \pi$. The simulation results confirm that the proposed ASMC achieves rapid trajectory tracking, with errors converging to zero within 0.5 seconds. Notably, there is no overshoot observed, indicating excellent stability and control precision. Figures 2 and 3 illustrate the trajectory responses and tracking errors of the joints, respectively, showcasing the controller's

5620...

ability to follow the desired trajectory accurately and efficiently.

- Uncertain Desired Trajectory

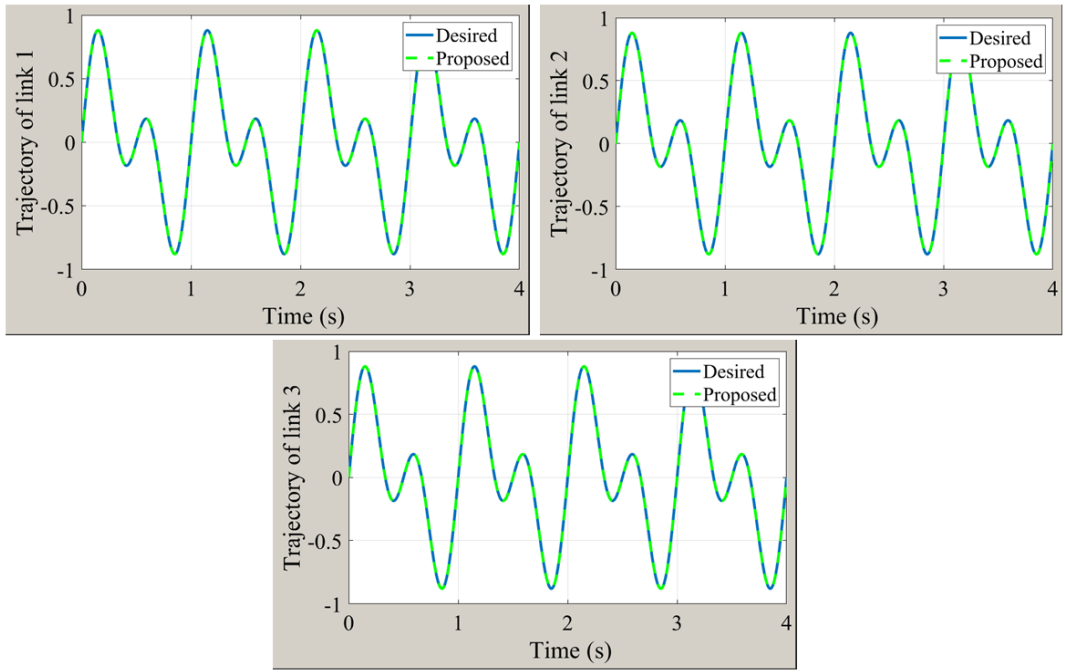

Figure 5. Trajectory responses of joints for the uncertain desired trajectory

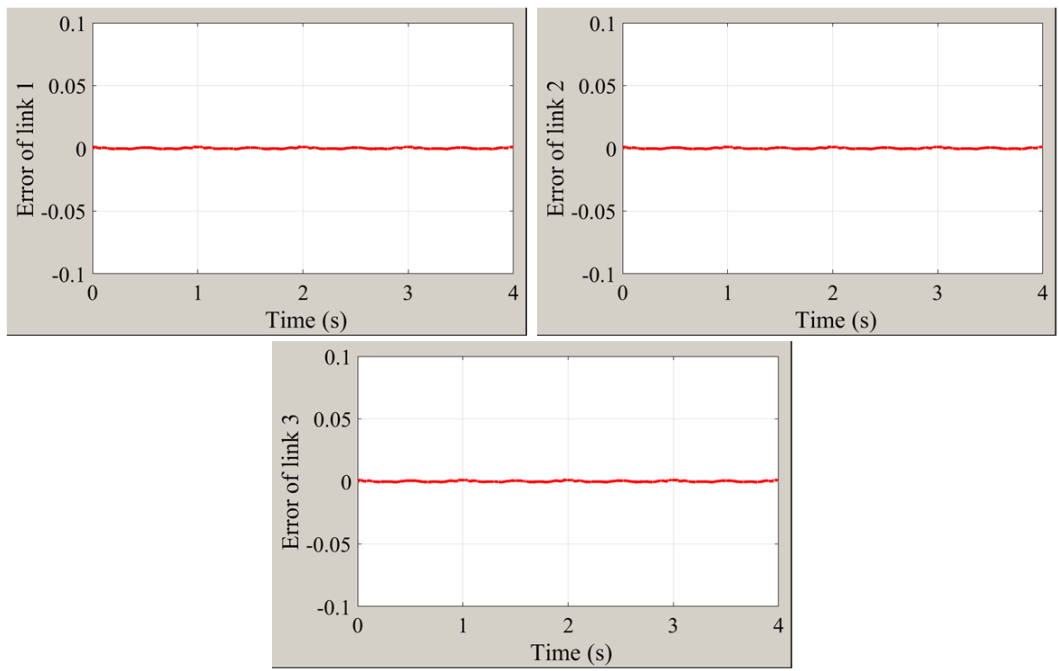

Figure 6. Tracking errors of joints for the uncertain desired trajectory

For an uncertain desired trajectory, the ASMC exhibits superior tracking performance. The controller effectively minimizes the tracking error while maintaining system stability. The results demonstrate smooth trajectory tracking without oscillations, even under varying conditions, highlighting the robustness and adaptability of the proposed method.

Figures 5 and 6 show the trajectory responses and tracking errors, confirming the high precision and stability of the ASMC under uncertain conditions.

- Sinusoidal Desired Trajectory

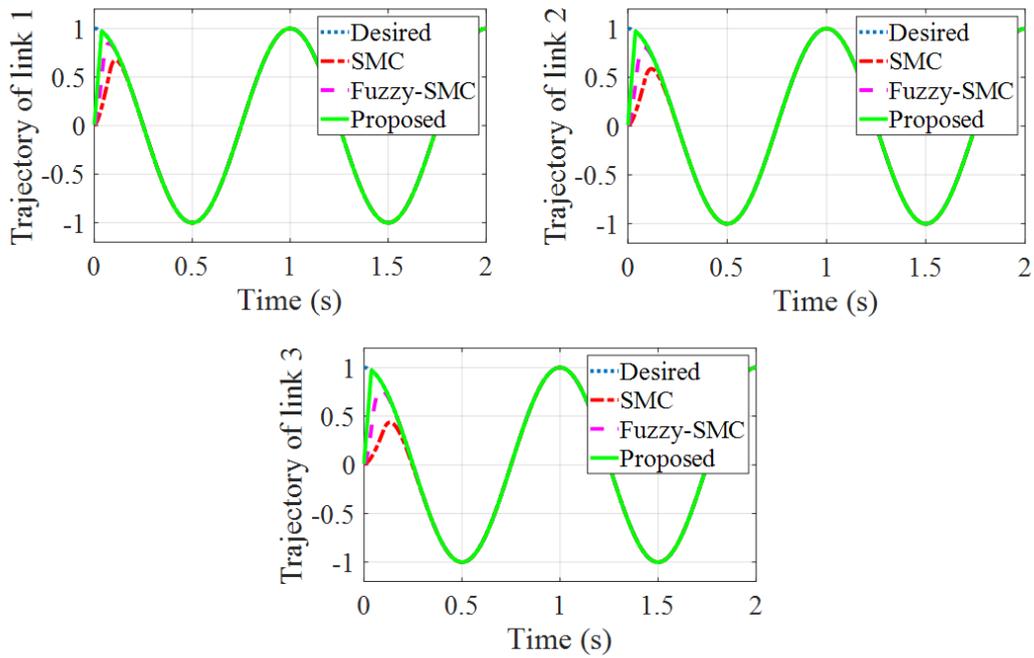

Figure 7. Trajectory responses of joints for the sinusoidal desired trajectory

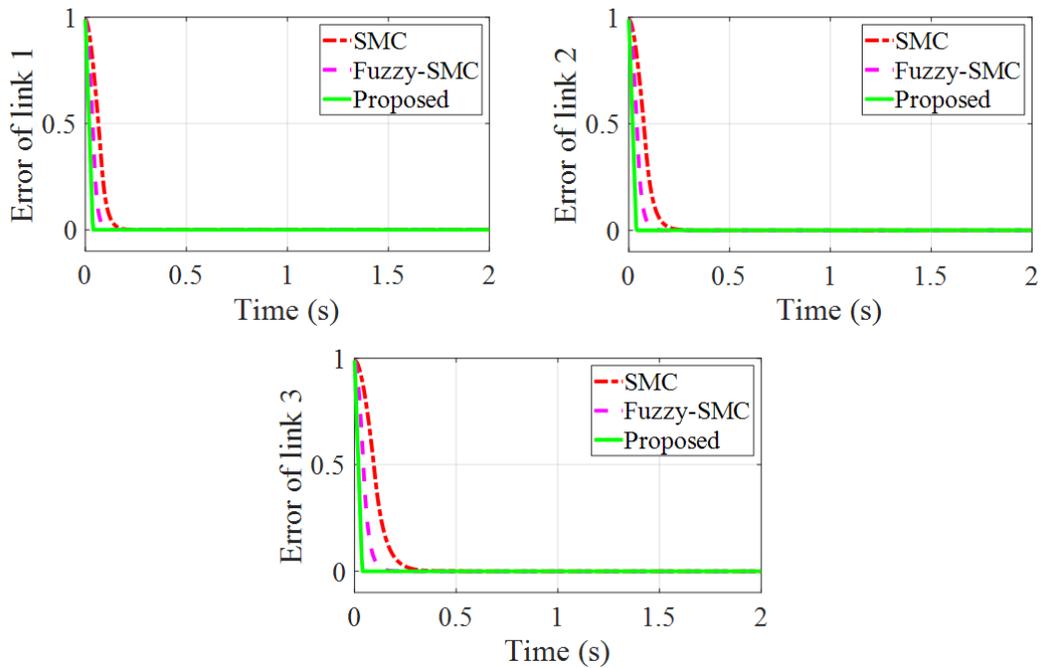

Figure 8. Tracking errors of joints for the sinusoidal desired trajectory

When subjected to a sinusoidal desired trajectory defined as $\theta_1 = q_2 = q_3 = 1\sin(2\pi t)$, the ASMC outperforms both the SMC and fuzzy-SMC controllers. The proposed controller achieves faster response times and lower tracking errors. The results underline the ASMC's superior ability to handle dynamic trajectories with high accuracy.

Figures 7 and 8 present the trajectory responses and tracking errors for the sinusoidal input, demonstrating the clear advantage of the ASMC over the comparison controllers.
- Disturbance at Joint 3

"Development of an Adaptive Sliding Mode Controller using Neural Networks for Trajectory Tracking of a Cylindrical Manipulator"

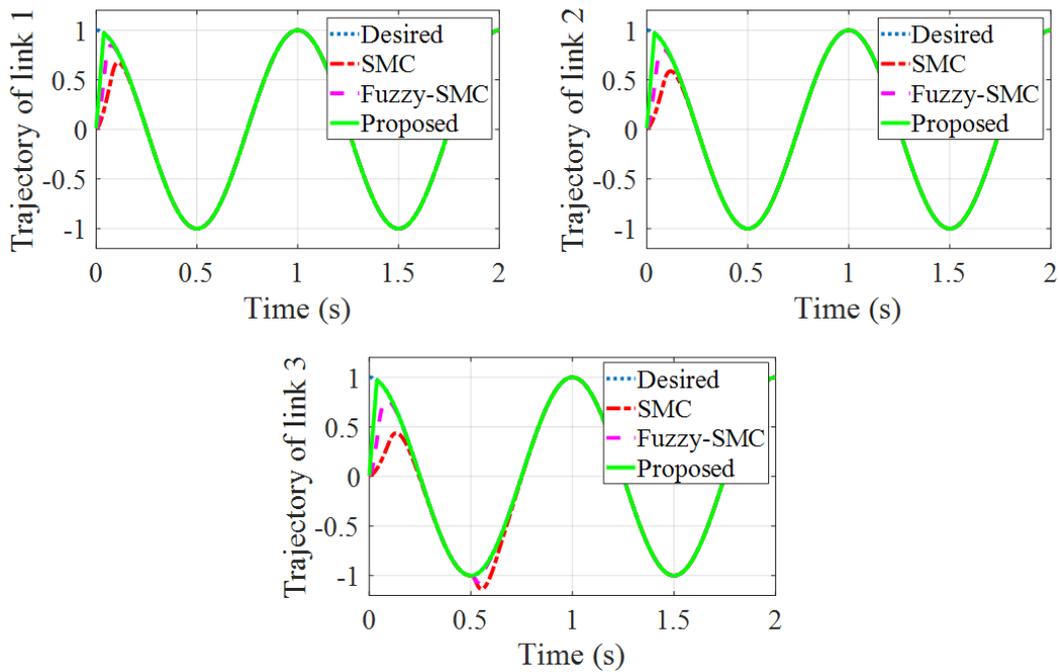

Figure 9. Trajectory responses of joints under disturbance at joint 3

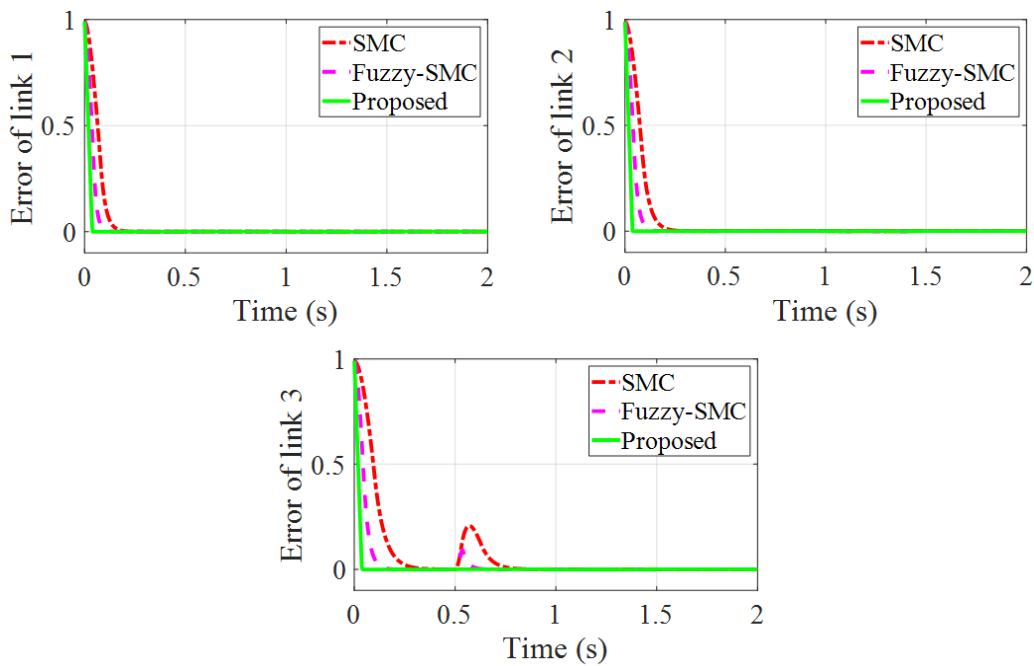

Figure 10. Tracking errors of joints under disturbance at joint 3

A disturbance was introduced at joint 3 at t = 0.5s The ASMC showed remarkable robustness by rapidly suppressing the disturbance and restoring the system to the desired trajectory. Compared to the SMC and fuzzy-SMC controllers, the ASMC demonstrated faster disturbance rejection and error recovery, underscoring its superior noise-handling capabilities.

Figures 9 and 10 depict the trajectory responses and tracking errors under this scenario, highlighting the resilience and effectiveness of the ASMC in maintaining performance under external disturbances.

V. CONCLUSION

This study presented the development of an Adaptive Sliding Mode Controller integrated with Neural Networks (ASMC-NN) to enhance trajectory tracking performance for cylindrical manipulators, particularly in applications like 3D printing. By combining the robustness of Sliding Mode Control with the adaptability of Neural Networks, the proposed controller effectively addressed system uncertainties and nonlinear dynamics, achieving high accuracy and reliability.

Simulation results demonstrated that the ASMC-NN significantly reduced tracking errors, improved response

5623

times, and enhanced system stability compared to traditional methods. These findings underscore the potential of the ASMC-NN approach for advancing control strategies in industrial automation and other precision-demanding applications. Future research may focus on experimental validations and extending the approach to more complex manipulator configurations.